\documentstyle[12pt]{article}
\newcommand{\gsim}{\mbox{\raisebox{-1.ex}{$\stackrel
{\textstyle >}
{\textstyle\sim}$}}}
\newcommand{\lsim}{\mbox{\raisebox{-1.ex}{$\stackrel
{\textstyle<}
{\textstyle \sim}$}}}
\newcommand{\square}{\kern1pt\vbox{\hrule height 1.2pt\hbox
{\vrule width1.2pt\hskip 3pt \vbox{\vskip
6pt}\hskip 3pt\vrule width 0.6pt}\hrule height 0.6pt}\kern1pt}
\newcommand{\vecF}{\!\!\mbox{ \boldmath $F$}}
\newcommand{\vecA}{\!\!\mbox{ \boldmath $A$}}
\newcommand{\vectau}{\!\!\mbox{ \boldmath $\tau$}}
\newcommand{\vecU}{\!\!\mbox{ \boldmath $U$}}
\newcommand{\vecPhi}{\!\!\mbox{  \boldmath  $\Phi$}}

\newcommand{\vecxi}{\!\!\mbox{ \boldmath $\xi$}}
\newcommand{\vecsigma}{\!\!\mbox{ \boldmath $\sigma$}}

\renewcommand{\Large}{\large}

\begin{document}

\begin{titlepage}

{}~\\

\vskip 0.3cm
\begin{center}
{\bf
\vskip 0.0cm
{\large \bf Non-Abelian Black Holes and Catastrophe Theory
\vskip 0.5cm
  I : Neutral Type
}

}\vskip .5in

{\sc  Takashi Torii }$^{(a)}$,
{\sc Kei-ichi Maeda}$^{(b)}$, and
 {\sc Takashi Tachizawa}$^{(c)}$
\\[1em]
 {\em Department of Physics, Waseda University,
Shinjuku-ku, Tokyo 169, Japan}

\vskip 1.2cm
{\bf Abstract}
\end{center}

\small
\baselineskip = 15pt
We   re-analyze  the globally neutral non-Abelian black holes
and present a unified picture, classifying them into
two types; Type I (black holes with massless non-Abelian field)
and Type II (black holes with ``massive" non-Abelian field).
For the Type II,
there are two branches:
The black hole in the high-entropy branch is ``stable" and almost
neutral,
 while that in
the  low entropy branch, which is similar to the Type I,  is
unstable and locally charged.
To analyze   their stabilities,
we adopt the catastrophe theoretic method, which reveals us a
universal picture
of stability of the black holes. It is shown that
the isolated  Type II black hole has  a fold catastrophe structure.
 In a  heat bath system, the Type I black hole
shows a cusp catastrophe,
while the Type II has  both fold and cusp catastrophe.

\vspace*{2mm}
 PACS numbers: 97.60.Lf, 11.15.-q, 95.30.Tg, 04.20.-q

\vspace{.2 cm}

\noindent
(a)~~electronic mail:
64l514@cfi.waseda.ac.jp\\
(b)~~electronic mail:
maeda@cfi.waseda.ac.jp\\
(c)~~electronic mail:
63l507@cfi.waseda.ac.jp
\end{titlepage}

\normalsize
\baselineskip = 18pt


\section{Introduction}

The discovery by Bartnik and McKinnon
of a non-trivial particle-like structure
(BM particle) in the Einstein-Yang-Mills system
is quite surprise\cite{Bar},
because it is known that there is no non-trivial solution in the
Einstein theory and Yang-Mills(YM) theory respectively and because
the work of Bekenstein\cite{Bek},
Hartle\cite{Har} and Teitelboim\cite{Tei} shows that stationary
black hole
solutions are similarly hairless in a variety of theories coupling
classical
fields to Einstein gravity.
After this solution, a variety of
self-gravitating structure with non-Abelian fields have been found.
Besides the BM particle, the colored black hole\cite{Vol} in the
same system,
the Skyrmion\cite{Sky,Dro} or the Skyrme black hole\cite
{Luc,Dro,Tor},
the dilatonic BM particle or dilatonic colored black hole\cite
{dil,Tor}
the particle solution with massive Proca field (Procaon) or the
Proca
black hole\cite{Gre}, the monopole\cite{'tH,LNW,Bre} or the black
hole in
monopole (monopole black hole)\cite{LNW,Bre,Aic}, and the
sphaleron\cite{Das,Gre}
or sphaleron black hole\cite{Gre} are discovered. Last two
solutions
 (monopole and sphaleron)
are obtained in Einstein-Yang-Mills-Higgs system.

However,
the colored black hole is unstable against the radial
perturbations\cite{Gal}. While, the Skyrme black hole solution
really
stirred our interest for the
non-Abelian black hole, since it turns out to be
a stable solution\cite{Heu}. Hence, it can be the likeliest
candidate of a counterexample for the black hole
no-hair conjecture\cite{Bizon}.
There are two Skyrme black holes with the same
horizon radius: one is stable and the other is unstable\cite{Tor}.
  The Proca black hole
and the sphaleron black
hole have the similar properties, although
 the latter
 is always unstable.
The monopole black hole is globally charged,  and stable  except for
that with fine-tuned parameters.
The dilatonic colored black hole is a natural
extension of that in Einstein-Maxwell-dilaton system,
which showed an interesting  feature in thermodynamics\cite{Gib}.
We found
that the dilatonic
colored black hole has similar but more complicated
properties\cite{Tor}.

Because of the non-Abelian gauge fields, each study
revealed that in spite of a simple ansatz (spherical symmetry)
those non-Abelian black holes
have complicated structures and show a variety of properties,
as shown in the above.
Are there any common properties in those
non-Abelian structures ? Can we find any universal understanding
for them?
Answering these questions is the main purpose of the present
paper.
In order to reach our goal,  we first classify  non-Abelian black
holes
into two classes; one is globally neutral ones and the other is
globally charged ones, because the charged monopole black
hole is obviously different from others and
should be discussed in a different context with
the Reissner-Nordstr\"om black hole.
We will discuss those two cases separately\cite{2-1}.  In this
paper we just
 focus on neutral case.
The neutral holes can be classified into two: one
is a class of black holes with  massless non-Abelian
field (Type I),
 and the other is that with ``massive"
non-Abelian field (Type II), i.e., the gauge or chiral field either has
a mass
in the original Lagrangian or get its mass through
a symmetry breaking mechanism by Higgs fields.
This classification provides us a unified picture
for non-Abelian black holes and their particle solutions.

One of the most important questions about the above  self-
gravitating
non-Abelian structure or black holes is , are they stable? We
answer
this question
via a catastrophe-theoretic analysis, which shows its power over
various phenomena.
Catastrophe theory may provide  us a clearer understanding than
a linear perturbation
method, because it visualizes the stability of non-Abelian black
holes.
Another merit in using this theory is that the way to investigate
the stability
becomes very
simple. The elementary catastrophe of Type I
black hole is found to be a fold type, which causes the existence of
stable
and unstable solutions. We also apply this to  stability analysis of
black holes in a heat bath
and show that more complicated high order elementary catastrophe
appears.
The swallows tail will appear in the monopole black hole\cite{2-
1}.
In \S 2, we shortly review non-Abelian black holes, and
provide a unified picture. The stability analysis of non-Abelian
black holes via a catastrophe
theory   are given
in \S 3. Section 4 includes
discussions and some remarks.


\vskip 2 cm
\setcounter{equation}{0}
\section{Summary of Non-Abelian Black Holes}

We have re-analyzed five
 non-Abelian theories, which are listed in Table 1.
Each theory has non-trivial self-gravitating
structure (particle solution) and non-trivial black holes.
We present a short
review on those black hole solutions and summarize their
properties
in order to find out a universal picture, which will be obtained in
the
end of this section.

\subsection{Colored Black Holes}

We first consider the Einstein-Yang-Mills system,
which  action is described as\cite{chu1};
\begin{equation}
  S = \int d^4x \sqrt{-g}\left[\frac1{2\kappa^2}R\left(g\right)
        -\frac1{16\pi g_{\rm C}^2}  \mbox{Tr} \vecF^2 \right],
       \label{2-1}
\end{equation}
where $\kappa^2=8\pi G$, and $\vecF$ is the YM field expressed by
its
potential $\vecA$ as
$\vecF =d\vecA +\vecA \wedge \vecA$.  $g_{\rm C}$ is a
self-coupling constant of the YM field.
In the spherically symmetric static
spacetime, the  metric is written as
\begin{equation}
  ds^2 = -\left(1-\frac{2Gm}r\right) \mbox{e}^{-2\delta}dt^2
         +\left(1-\frac{2Gm}r\right)^{-1}dr^2
         +r^2\left(d\theta^2+\sin^2\theta d\phi^2\right),  \label{2-2}
\end{equation}
where a mass function $m=m(r)$ and a lapse function
$\delta =\delta(r)$  depend only on
the radial coordinate $r$.
In order to find an asymptotically flat black hole solution with a
regular
event horizon,  we have to require the following boundary
conditions,

(i) asymptotic flatness, i.e.,  as $r \to \infty$,
\begin{eqnarray}
     m\left(r\right) & \to & M={\rm finite}      \label{2-3}   \\
     \delta \left(r\right) & \to & 0 .   \label{2-4}
\end{eqnarray}

(ii) the existence of a regular horizon $r_H$, i.e.,
\begin{eqnarray}
     2Gm_{\rm H}=r_{\rm H},    \label{2-5}  \\
     \delta_{\rm H}<\infty .  \label{2-6}
\end{eqnarray}

(iii) nonexistence of singularity outside the  horizon, i.e.,
for $r>r_{\rm H}$
\begin{equation}
     2Gm\left(r\right)<r ,  \label{2-7}
\end{equation}
where the variables with a subscript ${\rm H}$ denote those values
at the horizon.

The most generic form of a spherically symmetric SU(2) YM
potential\cite{Wit}
is
\begin{equation}
       \vecA =a\vectau_r dt+b\vectau_r dr
           +\left[ d\vectau_{\theta} -(1+w) \vectau_{\phi} \right] d
\theta
           +\left[ (1+w)\vectau_{\theta} +d\vectau_{\phi} \right]
           \sin \theta d \phi ,
       \label{2-8}
\end{equation}
where $a$, $b$, $d$ and $w$ are functions of  time and radial
coordinates, $t$ and  $r$. We have adopted the polar coordinate
description
$(\vectau_r, \vectau_{\theta}, \vectau_{\phi})$, i.e.
\begin{eqnarray}
     \vectau_r & = & \frac1{2i} [\vecsigma_1 \sin \theta \cos \phi
         + \vecsigma_2 \sin \theta \sin \phi
         + \vecsigma_3 \cos \theta] ,
        \label{2-9}  \\
     \vectau_{\theta} & = & \frac1{2i} [\vecsigma_1 \cos \theta
\cos \phi
         + \vecsigma_2 \cos \theta \sin \phi
         - \vecsigma_3  \sin \theta] ,
        \label{2-10}  \\
     \vectau_{\phi} & = & \frac1{2i} [-\vecsigma_1 \sin \phi
         + \vecsigma_2 \cos \phi] ,
        \label{2-11}
\end{eqnarray}
whose commutative relations are
\begin{equation}
     \left[ \vectau_a, \vectau_b \right] =\vectau_c
  \hspace{10mm} a, b, c, = r, \theta , ~{\rm or }~~\phi. \label{2-12}
\end{equation}
Here $\vecsigma_i \hspace{3mm} (i=1,2,3)$ denote the Pauli spin
matrices.
We can  set the ansatz $a \equiv 0$ ('tHooft ansatz, i.e.,
 purely magnetic YM field strength exists) and
eliminate  $b$ using a residual gauge freedom.
We can also set
$d \equiv 0$ in this case\cite{d_freedom}.
In the static case, the remaining function $w$ depend only on
the  radial coordinate $r$. As a result, we obtain a simplified
spherically symmetric YM
potential as
\begin{equation}
  \vecA = (1+w)[-\vectau_{\phi} d\theta +\vectau_{\theta}
  \sin \theta d\phi ].   \label{2-13}
\end{equation}
Substituting this  into $\vecF =d\vecA +\vecA \wedge \vecA$,
we have an explicit form of the field strength:
\begin{equation}
     \vecF   =  -w'\vectau_{\phi} dr\wedge d\theta
          +  w'\vectau_{\theta} dr\wedge \sin \theta d\phi
        -  (1-w^2)  \vectau_r d\theta \wedge \sin \theta d\phi.
        \label{2-14}
\end{equation}
As for the boundary conditions for YM field, we impose
\begin{equation}
 w \to \pm 1,   \label{2-15}
\end{equation}
which guarantees a finiteness of the energy of system.
On account of these boundary conditions (\ref{2-3}) $\sim$ (\ref
{2-7})
and (\ref{2-15}),
the basic field equations have to be
solved as an eigen value problem and the solutions are obtained
only by means of numerical analysis except for  trivial solutions
such as the Schwarzschild black hole
($w\equiv \pm 1$) or the Reissner-Nordstr\"om black hole
($w\equiv 0$).

A colored black hole has the following properties: First,
in the limit of $r_{\rm H} \rightarrow 0$, we find a particle-like
solution,
which  is  the
BM particle and  cannot exist without gravity.
For any values of mass, the colored black hole solutions exist, but
when its mass  increases, the structure gets similar to that of the
Schwarzschild black hole.  Because in the limit $r_{\rm H} \to
\infty$,
i.e.,  $M \to \infty$, the energy contribution from
the non-Abelian YM fields keeps still finite while the mass energy
of
the singularity at the center
must get to infinity.
The YM field makes no contribution to the black hole structure,
although
it can have nontrivial distribution on the Schwarzschild
background.
The colored black hole has
discrete mass spectrum, which is  characterized by the node
number of YM
potential. All solutions are unstable against the radial linear
perturbations.
The number of the unstable modes increases as the node number
 increases\cite{Wal}.

{}From the analysis of its temperature,  we showed that the sign
changes of the
specific heat occur twice\cite{Tor}. When the black hole is larger
some critical mass scale ($M_{\rm 1, cr} = 0.905m_{\rm P}/g_{\rm
C}$) or smaller
than the other critical mass scale ($M_{\rm 2, cr} = 1.061m_{\rm
P}/g_{\rm C}$), the
specific heat is negative, but in
between two critical mass scales,  the effective charge, which is
defined by a
 surface integration of the YM field at the horizon, becomes so
large that
the YM field
seems to be dominated and consequently its specific heat becomes
positive.

\subsection{Dilatonic Colored Black Holes}

Next, we consider models with the following action;
\begin{equation}
  S = \int d^4x \sqrt{-g}\left[\frac1{2\kappa^2}R\left(g\right)
        -\frac1{2\kappa^2}\left( \nabla \sigma\right)^2
        -\frac1{16\pi g_{\rm C}^2} \mbox{e}^{-\alpha \sigma}
         \mbox{Tr} \vecF^2 \right],
       \label{2-16}
\end{equation}
where $\sigma$ is a dilaton field, and $\vecF$ and $\vecA$ are the
YM field
strength  and its potential, respectively.
$\alpha(\geq 0)$ is a coupling constant of the dilaton field
$\sigma$ to
the YM field $\vecF$.
This type of action arises from various unified theories including a
superstring
model\cite{Gib}. For example, $\alpha=1$ and
$\alpha=\sqrt{3}$ are the cases arising from a superstring theory
and from the 5-dimensional Kaluza-Klein theory, respectively.
Setting $\alpha =0$ with
$\sigma \equiv 0$,
the model (\ref{2-16}) reduces to the Einstein-Yang-Mills system,
which
was discussed in \S 2.1.

The dilatonic colored black hole is an extension of the colored
black
hole, i.e.,  the colored black hole with a scale
invariant massless scalar field $\sigma$. The spacetime and the
YM potential ansatz
are the same as those of the colored black hole. As for  the
asymptotic
behavior of dilaton field $\sigma$, we impose
\begin{equation}
\sigma \to 0  ,
\hspace{10mm} \mbox{as} \hspace{5mm} r\to \infty , \label{2-17}
\end{equation}
in order to make the energy of  system finite.

The properties of a dilatonic colored black hole are quite similar to
the colored black hole besides its thermodynamical behavior. For a
small
coupling constant model  ($\alpha <0.5$)  the sign change of
specific
heat occurs twice as the  colored black hole does. However,  for the
larger
the coupling constant $\alpha$, there is no sign change.
The similar aspect was also seen for the
Einstein-Maxwell-dilaton (EMD) system, although the criteria for
the sign  change of specific heat was somewhat different, i.e.,
$\alpha <1$\cite{Gib}.
It is remarkable that the extension of a gauge field such as  U(1)
$\to$
SU(2) preserve the similar  property. This may be caused by the
fact that
the (dilatonic) colored black hole has the
similar spacetime structure to that of Reissner-Nordstr\"om black
hole
near the horizon and the strong coupling reduces its effective
charge as in
the EMD system.

\subsection{Proca Black Holes}

Next we consider a particle-like structure with massive
``YM" field (Proca field)
and its
black hole solutions.
Those are  obtained from the action
\begin{equation}
  S = \int d^4x \sqrt{-g}\left[\frac1{2\kappa^2}R\left(g\right)
        -\frac1{4\pi g_{\rm C}^2} \left( {1 \over 4} \mbox{Tr} \vecF^
2
        - {\mu ^2 \over 2} \mbox{Tr}  \vecA^2 \right)
        \right],
       \label{2-24}
\end{equation}
where $\mu$ is a mass of the vector field $\vecF$.
Although the mass term breaks a gauge invariance
and such a theory is not renormalizable,
the Proca model may be useful as an  effective theory of massive
spin 1
field. Furthermore, it may also be a  simple and good model to
understand
 common properties of black holes in Type II black hole which
includes more
realistic models discussed in the next two subsections,
and to reveal its essence.

We assume a spherically symmetric static metric (\ref{2-2})
and  vector potential (\ref{2-13}) and that their boundary
conditions
are the same as those of  colored black hole.

Solving the Einstein-Proca equations, we show the $M-r_{\rm H}$
relation
in Fig. 1(a).
The  Proca black holes have the
following properties: In the limit of $r_{\rm H} \rightarrow 0$, we
find
 two particle-like solutions. One corresponds
to a self-gravitating particle solution (we shall call it Procaon),
which can exist without  gravity, i.e., in the Minkowski space,
and the other has similar properties  to
those of the BM particle. Two branches of black hole
solutions, which leave from those two particles, merged at some
critical
horizon
radius. Beyond this critical point, where the black hole has a
maximal mass $M_{\rm C}$ and a maximal horizon radius $r_{H,
{\rm C}}$,
there exist no non-trivial structure.
One interesting feature in Fig. 1(a), which turns out to be
important later,
is that there is a cusp structure at the critical point.
Since the black hole entropy is given by
\begin{equation}
S=A/4=\pi r_{\rm H}^2 ,
\end{equation}
 where $A$
is its area, the
upper branch in Fig. 1(a) has larger entropy than that in the lower
branch.
Hence, we shall call each of them, the high- and low-entropy
branches
respectively. The low-entropy branch is similar to the colored
black hole
solution, and in fact it approaches
to the colored black hole in the ``low-energy" limit.
The high-entropy branch approaches the Schwarzschild black hole
in the ``low-energy" limit. Here, the ``low-energy" limit means that
the mass of the non-Abelian field  $\mu$
 is much smaller than the Planck mass, $m_{\rm P} =
G^{-\frac12}$.
In the limit of ``high-energy", no solution exist. Both branches
vanish around
$\mu \sim m_{\rm P} /g_{\rm C}$.

{}From the stability analysis by a catastrophe theory given in the
next section,
we expect that
the high-entropy branch is stable, while
the low-entropy branch is unstable  against  radial
perturbations. Since
this  was numerically proved for the Skyrme black holes (see \S
2.4),
we believe that it is also true
for the Proca black holes from
our unified picture.

What makes the difference between the high-entropy and the
low-entropy branches? To understand it from a stability
point of vies, we shall see the energy distribution of the YM
field. The energy of YM fields consists of two terms, $\rho_
{\rm F^2}$ which is from the kinetic term and $\rho_{\rm A^
2}$ which is from the mass term. In Fig. 2 we show $\rho_{\rm
F^2}$ and $\rho_{\rm A^2}$ of the Proca black hole
separately. The total energy density is the sum of them, i.e.
$\rho_{\rm total} = \rho_{\rm F^2}+\rho_{\rm A^2}$. $\rho_
{\rm A^2}$ decays as $r^{-2}$ near the horizon because $w
\sim 1$ and it drops very quickly at some radius around the
Compton wave length of YM field $(\sim 1/\mu)$ which is
marked by an arrow in Fig. 2. As for $\rho_{\rm F^2}$, it is
complicated because various factors are tangles. But it also
decay rapidly at the same radius as $\rho_{\rm A^2}$ does
so. The main difference between the high-entropy and low
entropy branches is that $\rho_{\rm A^2}$ is still dominated
at the Compton wave length in the high-entropy branch while
$\rho_{\rm F^2}$ becomes dominant there in the low-entropy
branch. We in general expect the mass term give a
contribution to stabilize the structure. This is the case for
the high-entropy branch because the typical size of the
structure is about $1/\mu$, where $\rho_{\rm A^2}$ is still
dominant. On the other hand, in the low-entropy branch
$\rho_{\rm F^2}$ becomes already dominant inside of the
structure, then the mass term cannot stabilize the system.
The extreme case is the colored black hole $(\rho_{\rm A^2}
\equiv 0)$.

The specific heat in the high-entropy branch is always negative
like the
Schwarzschild black hole, while the specific heat in the low-
entropy branch
change its sign a few times.
We show a numerical result in Fig. 3(a).
When the mass of the non-Abelian field
 is small, i.e., in the  ``low-energy" case, we find the sign
 changes three times which occur at the
critical mass scales of  $M_{\rm 1,cr}, M_{\rm 2,cr},$ and
$ M_{\rm 3,cr}$ (In Fig. 3(a), $M_{\rm
1,cr}=0.909m_{\rm P}/g_{\rm C},  M_{\rm 2,cr}=1.06m_{\rm P}/g_
{\rm C},
 {\rm and}~
 M_{\rm 3,cr}=1.62m_{\rm P}/g_{\rm C}$, for $\mu =0.05m_{\rm
P}/g_{\rm C}$).
 The critical point $M_{\rm 2,cr}$
is the same type as that of the Reissner-Nordstr\"om black hole
and the
smaller one ($M_{\rm 1,cr}$) corresponds to the lower critical
point in the
(dilatonic) colored black hole type. Between $M_{\rm 1,cr}$ and
$M_{\rm
2,cr}$, the specific heat becomes positive, when
the effective charge at horizon becomes large. The larger critical
point
 ($M_{\rm 3,cr}$) appears only in Type II
 black holes.
 Beyond $M_{\rm 3,cr}$, the specific
heat becomes positive again. As the mass of the non-Abelian field
$\mu$
increases, however,  $M_{\rm 1,cr}$ and $M_{\rm 2,cr}$ merge and
the sign
change of the specific
heat changes occurs only once. This change may describe
some important feature for
 the stability of black holes in a heat bath (see \S 3.2).

\subsection{Skyrme Black Holes}

In the following two subsections,  we consider two realistic
models.
Here we deal with the Skyrme field as non-Abelian fields.
The SU(2)$\times$SU(2) invariant action coupled to gravity is
given by
\begin{equation}
  S = \int d^4x \sqrt{-g}\left[\frac1{2\kappa^2}R\left(g\right)
        +\frac14 f_{\rm S}^2 \mbox{Tr} \vecA^2-\frac1{32 g_{\rm S}^
2}
        \mbox{Tr} \vecF^2 \right],
\label{2-18}
\end{equation}
where $\vecF$ and $\vecA$ are the field strength and its potential,
respectively, and
$f_{\rm S}$ and $g_{\rm S}$ are coupling constants.
We use the gauge coupling constant $g_{\rm S}$ instead of
 $g_{\rm C}$, both of which are  the same
up to a constant, i.e.,
\begin{equation}
g_{\rm S} = \sqrt{4\pi} g_{\rm C} .  \label{2-19}
\end{equation}

For a spherically  symmetric static solution,
the spacetime metric is
 (\ref{2-2}). As for the boundary conditions,  we
again require the asymptotic flatness (\ref{2-3}), (\ref{2-4}),
the existence of a regular event horizon (\ref{2-5}),
 (\ref{2-6}),
 and the regularity of the
spacetime (\ref{2-7}).

$\vecA$ and $\vecF$ are described
in terms of the SU(2)-valued function $\vecU$ as
\begin{equation}
  \vecA=\vecU^{\dag} \nabla \vecU ,\;\;\;\; \vecF=\vecA\wedge
\vecA.  \label{2-20}
\end{equation}
 In a spherically symmetric static case, we make the hedgehog
ansatz for
$\vecU$, i.e.,
\begin{equation}
  \vecU\left(x\right)=\cos\chi\left(r\right)+i\sin\chi\left
(r\right)
      \mbox{\boldmath $\sigma_i $}
      \hat{r}^i,  \label{2-21}
\end{equation}
where the $\vecsigma_i$ denote the Pauli spin matrices and $\hat
{r}^i$ is a
radial normal. The finiteness of the energy of  system yields the
asymptotic boundary condition for $\chi$ as;
\begin{equation}
  \chi \to 0, \hspace{10mm} \mbox{as}\hspace{5mm} r \to \infty .
\label{2-22}
\end{equation}
For the particle-like solution (Skyrmion), the value of $\chi$ at the
origin
must be
$2\pi n$,
where $n$ is an integer and $|n|$ denotes the winding number of the
Skyrmion.
But in the case of the black hole solution, because it is
topologically
trivial, the winding number  defined by
\begin{equation}
  W_n\equiv \frac1{2\pi} \left|\chi_{\rm H}
        -\sin(\chi_{\rm H})\right| ,
   \label{2-23}
\end{equation}
is no longer an integer\cite{Luc}. Nevertheless, since $W_n$ is
close to $n$,
we shall also call $n$ the ``winding" number of the Skyrme black
hole.

The properties of Skyrme
black holes are quite similar to those of  the Proca black holes.
Replacing the mass parameter $\mu$ in the Proca model with
that in the Skyrme model, i.e., $\mu = g_{\rm S} f_{\rm S}$,
all discussions in \S 2.3 is applied to the  Skyrme black hole
except for a  difference between symmetries of their vector fields.
It is worth noting  that a family of solutions provides a
cusp structure in the $M-r_{\rm H}$ plane
as the Proca black holes (see Fig. 1(b)).
As we already mentioned in the Proca model,
the high-entropy branch is stable, while
the low-entropy branch is unstable\cite{Dro,Luc,Gal,Heu} against
radial
perturbations
and this property is understood by a catastrophe theory.
The specific heat is also quite similar to that of the Proca black
holes
(see Fig. 3(b)).

In addition, we should remark the following
argument derived from its topological structure in the Skyrme
model.
For each exited state, i.e., each solution with  larger winding
number,
we find the same cusp structure, which consists
of the high- and low- entropy branch,
in the $M -r_{\rm H}$ plane\cite{Tor}.
Since the colored black hole has $n$ unstable modes for an $n$-
node
solution\cite{Gal,Wal},  we expect that the low-entropy branch
of the Skyrme
black holes with the
winding number  $n$, has $n$
unstable modes.
It is confirmed by the fact that the low-entropy branch approaches
the colored black hole in the ``low-energy" limit.
While,
the high-entropy branch may have no unstable mode, because
the particle-like Skyrmion solution is topologically non-trivial,
 i.e., its homotopy group is not trivial.
Although its black hole solution does not have such
a topological invariance,
they may be  stable.
There is no proof, but we believe that there
is no stability change between a particle-like solution and
 the associated black hole solutions. In fact, we will show
 in the following section that  a
 catastrophe theory can be used to analyze the stability.
Then we expect that  there is no stability change
between a particle-like solution and the associated black hole
solutions and
the Skyrme black hole solutions
in high-entropy branch with any winding number must be stable.

\subsection{Sphaleron Black Holes}

As a final example,
we start with the following bosonic part of the Lagrangian in
the standard theory:
\begin{equation}
  S = \int d^4x \sqrt{-g}\left[\frac1{2\kappa^2}R\left(g\right)
        -\frac1{16\pi g_{\rm C}^2} \mbox{Tr} \vecF^2
        -\frac1{4\pi } \left(D_{\mu} \vecPhi \right)^{\dag}
            \left(D^{\mu} \vecPhi \right)
        -\frac1{4\pi } V\left( \vecPhi \right)
        \right],
       \label{2-25}
\end{equation}
where
\begin{eqnarray}
     D_{\mu} & = & \partial_{\mu} + \vectau \cdot \vecA_{\mu} ,
                            \label{2-26}    \\
     \vecF & = & d\vecA + \vecA \wedge \vecA ,
                            \label{2-27}  \\
     V\left( \vecPhi \right) & = & \lambda (\vecPhi^{\dag} \vecPhi -
\Phi_0^2)^2.
                           \label{2-28}
\end{eqnarray}
$F$ and $A$ are the SU(2) YM field strength and its potential,
respectively,
and $\vecPhi$ is the Higgs field. $g_{\rm C}$ is a coupling constant
of the YM field, and $\lambda$ and $\Phi_0$ are a self-coupling
constant
and a vacuum expectation value of the Higgs field, respectively.

The most general complex doublet of Higgs fields can be written as
\begin{equation}
              \vecPhi (x) = \frac1{\sqrt{2}}
                 \exp \left[ -\vectau \cdot \vecxi (x) \right]
              \left( \begin{array}{c} 0 , \\  \Phi(x)/r \end{array} \right)
{}.
       \label{2-29}
\end{equation}
For spherically symmetric static solution we impose the usual
ansatz\cite{chunsb};
\begin{equation}
\vecxi (x) =2\pi \hat{\mbox{\boldmath $r$}},
\hspace{10mm} \Phi(x) =\Phi (r) ,        \label{2-30}
\end{equation}
The asymptotic behavior of the Higgs field is assumed to be
\begin{equation}
\Phi(r) \to \Phi_0, \hspace{10mm} {\rm as} \hspace{5mm} r \to
\infty ,
 \label{2-31}
\end{equation}
which means that the vacuum is in a symmetry broken state at
infinity.
As for the spacetime metric, the YM potential, and the boundary
conditions
 we again set (\ref{2-2}),
(\ref{2-8}), (\ref{2-3})$\sim$(\ref{2-7})
and (\ref{2-15}).

Using the above ansatz we solve the Einstein-Yang-Mills-Higgs
equations numerically
by means of shooting method.
The shape of a family of solutions in the $M-r_{\rm H}$
plane (Fig.1(c)) and the behavior of the specific heat (Fig. 3(c))
 may lead  us to believe that the sphaleron black holes
 have the same properties as the Skyrme black hole and the Proca
black hole.
The reason why we find the same properties is that the  gauge field
gets its  mass,  $\mu = g \Phi _0$, through a spontaneous
symmetry
breaking by the  Higgs field.
However,  one important  difference exists from the point of view
of its
stability.
For the Skyrme black hole and the
Proca black hole, high-entropy branch is stable against radial
linear perturbations,
but the sphaleron black
hole is always unstable because of the topological reason.
 When we discuss stability in
general, there are many modes to be investigated. A general
argument about the
instability of the sphaleron is
 based on a topological analysis\cite{Das}, which
does not specify any  modes. For the sphaleron without gravity, the
stability analysis  with a spherically symmetric ansatz was
done\cite{Aki}. It
was explicitly shown that there is only one unstable mode. For the
case of
self-gravitation sphaleron or the sphaleron black hole, it is stable
in the
high-entropy branch against radial perturbations except for one
unstable
mode corresponding to the above. In the low-entropy branch,
at least one more unstable mode appears\cite{Str}.
In this sense, the high entropy branch is more stable than
the low-entropy branch.
Hence, we argue that the high-entropy
branch is ``stable" while the low-entropy one is unstable. With
this argument, we can
make a unified picture for all neutral cases including
the sphaleron black hole as well.

\subsection{A Unified Picture}

{}From the above summary,
we can classify non-Abelian  black holes by their properties
into two types: Type I (black holes with
massless non-Abelian field) and Type II (black holes with
``massive" non-Abelian field).
The colored
black hole and the dilatonic colored black hole
are classified as Type I. While the Proca black hole,
the Skyrme black hole,
 and the sphaleron black hole are Type II.
Those non-Abelian
fields
have a ``mass"  $\mu$.
$\mu = g_{\rm S} f_{\rm S}$
 for the
Einstein-Skyrme system, and  $\mu = g_{\rm C}\Phi_0$
 for the Einstein-Yang-Mills-Higgs system.
Each type of black hole has  the following common properties.

\vskip .5cm
\noindent
\underline{1. Type I Black Hole}\\
(1) This type of black hole is in an equilibrium state
by   balance between  a repulsive force by
the massless gauge field and the gravitational force.
It has  no upper bound of
their mass.
It may be essential for no upper bound of a mass that
the non-Abelian field is massless.
When the mass gets large, however, the spacetime approaches to
the Schwarzschild black hole.\\
(2) The field strength at the horizon
 $B_{\rm H}$, which is defined by
\begin{equation}
B_{\rm H} \equiv \left. ({\rm Tr} \vecF^2)^{\frac 12} \right|_{\rm
horizon} ,
\end{equation}
may give us a naive idea
about how much the black hole is locally  charged.
Here, the expression $B_{\rm H}$ has been used because only the
radial
component of magnetic part of non-Abelian field is finite at the
horizon.
{}From Fig. 4, in which we show $B_{\rm H}$ with respect to $M$, we
find that
the colored  black hole is locally charged.
Although they are  globally neutral, their structure near the
horizon
is similar to that of the Reissner-Nordstr\"om black hole.
When the black hole gets large, $B_{\rm H}$ decreases, that is, it
becomes
effectively less charged because it approaches the Schwarzschild
solution.
\\
 (3) These black holes are unstable against radial perturbations.
The number of unstable modes increases as the node number gets
large.\\
(4) The sign changes of the
specific heat occur twice, but
 for the large
the coupling constant ($\alpha \gsim 0.5)$ in the dilatonic colored
black holes,
 there is no sign change.
This is because  the strong coupling reduces its effective charge
 near the horizon.  Those behaviors can be understood by
a stability analysis of the black holes in a heat bath via
a catastrophe theory (see \S 3.2).\\

\vskip .5cm
\noindent
\underline{2. Type II Black Hole}\\
(1) This type of black hole has  an upper bound of the
mass scale, beyond which no non-trivial solution exist.
There are two branches: one is
``stable" and the other is unstable.
In the ``low energy" limit ($\mu \rightarrow 0$),
the stable branch approaches the Schwarzschild
black hole, while the unstable one converges to the colored black
hole.
The typical mass scale of non-trivial particle in the unstable
branch is about
$m_{\rm P}/g_{\rm C}$.

We suspect that the mass of the non-Abelian field causes the
existence of the upper bound of the black hole mass, though it has
not been
proved.
The non-trivial structure balanced between a repulsive force by
the massive gauge field and the gravitational force
may have a characteristic size $1/\mu$, which is the Compton
wave length.
The horizon of non-Abelian black hole must exist inside of the
structure,
otherwise the black hole swallows whole non-trivial structure,
resulting
in the trivial Schwarzschild black hole. The condition of
$r_{\rm H} \lsim 1/\mu$ yields the upper bound of the black hole
mass.

Furthermore, when  a mass of the non-Abelian field $\mu$
is  large enough,
the horizon scale $(\lsim m_{\rm P}/g_{\rm C})$ becomes larger
than $1/\mu$,
and then the
non-Abelian field is again swallowed into the black
hole beyond the event horizon. For this reason, there is no
non-trivial solution in the
``high-energy" limit ($\mu \sim m_{\rm P}$). \\
(2) The high-entropy branch is ``stable", while the low-entropy
branch
 is unstable. This behavior of stability is well understood by a
catastrophe theory (see \S 3.1).
\\
(3) As for the structure of Type II black hole, those in the low-
entropy branch is quite similar to the Type I. However, the
structure of black holes in the high-entropy branch may be
different from that of Type I. In the high-entropy branch,
$\rho_{\rm A^2}$ is always dominant in the whole structure,
which exist effectively until the Compton wave length $\sim
1/\mu$. It stabilizes the black hole. On the other hand in the
low-entropy branch $\rho_{\rm F^2}$ becomes dominant
inside of the structure. This is one of the reasons why the
low-entropy branch is still unstable as the colored black hole.
\\
(4) The specific heat in the high-entropy branch is always negative
like the
Schwarzschild black hole, while that in the low-entropy branch
changes its sign
 a few times depending on the mass of the non-Abelian field $\mu$.
When  $\mu$
 is small, i.e., in the  ``low-energy" case, we find the sign
 changes three times at $M_{\rm 1,cr}, M_{\rm 2,cr}, {\rm and }
 M_{\rm 3,cr}$.
The second critical point ($M_{\rm 2,cr}$)
is the same type as that of the Reissner-Nordstr\"om black hole
and the
lowest one ($M_{\rm 1,cr}$) corresponds to the lower critical
point in the
Type I black hole. Between $M_{\rm 1,cr}$ and
$M_{\rm 2,cr}$, the specific heat becomes positive.
 Beyond the largest critical point $M_{\rm 3,cr}$, which  appears
only in
Type II black holes, the specific
heat becomes positive again. But as $\mu$
increases, $M_{\rm 1,cr}$ and $M_{\rm 2,cr}$ merge and the sign
change of the specific
heat changes occurs only once. These changes may describe
some important feature for
 the stability of black holes in a heat bath, which can be
analyzed by a catastrophe theory (see \S 3.2).\\
(5)
$B_{\rm H}$
is still small for the high-entropy branch.
This black hole is approximately neutral. On the other hand,
for the low-entropy branch, $B_{\rm H}$  is finite and rather large
as seen from Fig. 4.  This type of  black hole
is locally charged but globally
neutral just as the colored black hole.\\

The negative specific heat in the high-entropy branch
is also consistent with that of the Schwarzschild
or Schwarzschild-de Sitter spacetime. On the other hand,
 for low-entropy branch,
 although the black hole is globally neutral, $B_{\rm H}$ does not
vanish at the horizon and the black hole is locally charged, which
yields
the sign change of the specific heat.

We summarize the above properties in Table 2.


\vskip 2 cm
\setcounter{equation}{0}
\section{Stability of Black Holes via Catastrophe Theory}

As we mentioned, Type I  black hole is
unstable against radial perturbations for any coupling constant or
in any mass
scale, but in the case of Type II its aspect is rather
complicated. They have two types of solutions; one of which is
``stable" and
the other is unstable, and they merge at a critical mass
$M_{\rm C}$ in the $M-r_{\rm H}$
plane with a cusp. There are various phenomena in nature
whose stabilities change at this
kind of cusp, and such a  stability change is understood
 by means of a catastrophe theory.

The catastrophe theory is a new mathematical tool to explain a
variety of change of  states in  nature, in particular
a discontinuous change of  states
which occur eventually in spite of gradual changes of parameters
of a system. It is widely applied
 today in various research fields, e.g., the  structural stability,
the crystal lattice, biology, embryology, linguistics etc.,  and as a
matter of
course, in  astrophysics\cite{Kus,Kab}. Here we shall
 apply the catastrophe
theory, which is different from a usual linear perturbation method,
to
analyze a stability of
non-Abelian black holes.

In order to examine a stability with the catastrophe theory, we
must first find control parameters, state variables and a potential
function
of the non-Abelian black hole system. They could be different
depending on the type of the
black hole (whether Type I or II) and on the
environment
around the black hole (whether the isolated system or that in a
heat bath).

The next step to examine a stability of a certain system with the
catastrophe
theory is to draw  the equilibrium space $M_V$, which consists of
extrema of the potential function, i.e., of a family of solutions of
the system,
in the  space of control parameters and state variables. Then we
project
the equilibrium space onto a control parameter space (which may
also
be called the control plane
in the 2-dimensional case as the present models) by
 a catastrophe map $\chi_V : M_V \to R^N$, where $N$ is the
number
 of the control parameters. There may exist singular points on
the equilibrium space, where the Jacobian of the mapping $\chi_V$
 vanishes.  The image of the set of  singular points
 is called the bifurcation set $B_V$.
If such singular points exist, the mapping $\chi_V$
from $M_V$ to  the control parameter
space is singular and then the number of solutions with the same
control parameters changes beyond the bifurcation set $B_V$
and then the stability also changes there.
Hence, by looking at this bifurcation set, we can classify
our models  into several
elementary catastrophe  and show the properties about
stability of the system as will be shown in below\cite{Tho}.

We have summarized our results in Table 3
together with the types of elementary catastrophe.

\subsection{Isolated Black Holes}
\subsubsection{Type I non-Abelian black hole}

Assuming that the state variable is the field strength at horizon
$B_{\rm H}$, the control parameters
are  the mass of black hole
$M$ and the coupling constant $\alpha$, and the potential function
is
the black hole entropy $S=\pi r_{\rm H}^2$ ,
we performed the above procedure for the dilatonic colored black
hole.
We find that its bifurcation set is an
empty set, namely, all points in the equilibrium space are
regular.  Hence, the dilatonic colored black
hole has no change of stability in any mass scale or for any
coupling constant.
This conclusion is consistent with the known fact that all
dilatonic colored
 black holes are unstable, which is obtained from the linear
perturbation method.

\subsubsection{Type II non-Abelian black hole}

Next we  examine the Type II black hole, which is more
interesting than Type I, because the cusp in
$M-r_{\rm H}$ plane provides us some catastrophe theoretic
features. We show
the equilibrium space and the bifurcation set of the Skyrme black
holes
in Fig. 5.
We have here adopted $M$ and the mass of non-Abelian field
$\mu(=g_{\rm S}f_{\rm S})$
as control parameters, and
$B_{\rm H}$ as a state variable,
and $S$ as a potential function, respectively.
The equilibrium  space looks
like a piece of cloth folded and the catastrophe mapping from the
equilibrium space to the control plane provides a line,
which is the evidence of a fold catastrophe.

When a system has two control
parameters like in the present case, two types of catastrophe are
possible;
one of which is
a cusp catastrophe and the other is a fold catastrophe.
To explain what will happen,
however, the latter does not need two control parameters but only
one  parameter.
Hence we have only to focus one parameter in the present model.
We  keep our eyes on the change of black hole mass $M$ leaving the
mass of
non-Abelian field
$\mu$ fixed. There is two reasons
 why we
choose $M$ as the essential control parameter instead of $\mu$.

One is that the fundamental parameters in the Lagrangian (the
mass of the particle $\mu$,
the coupling constants $g_{\rm C}$, $g_{\rm S}$, $f_{\rm S}$,  the
vacuum expectation
value $\Phi_0$ and so on) have certain fixed values for each theory
and then it
may be  physically
meaningless, though mathematically
interesting, to consider a change of the mass of non-Abelian field
$\mu$\cite{comment}.
Another reason is
that we are interested in what will happen on  such a black hole,
 in particular, on
how does its stability change
when the mass of black hole changes through accretion of
surrounded matter
or by the Hawking radiation.

Fixing one control parameter $\mu$, we find a smooth solution
curve on the equilibrium space (which is called
the equilibrium curve).
There exist two solutions  on this curve for each mass. The
solution
with the smaller value of $B_{\rm H}$  is stable
against radial perturbations,  i.e., in the high-entropy branch, while
that with larger value of   $B_{\rm H}$ is
unstable, i.e., in the low-entropy branch.

How does  a catastrophe come  in this system?  Assume that there
is a
stable black hole solution expressed by some point on the curve
labeled $e$ in
Fig.5. When its mass gets large, the point shifts to the
right along the curve $e$,
and  reaches eventually
to the end point (a singular point), beyond which there is no
solution. Then the solution point is forced to jump to
other stable solution
discontinuously as a solid arrow in Fig.5. This is a catastrophe.
This phenomenon
is explained by watching a potential function.
The schematic forms of the potential function (the black hole
entropy $S$)
depend on the control parameters as
shown  in Fig.6. In the region of small $\mu$  and small $M$
(the left hand side of $B_V$),
the potential function has two extremal points, one of which is a
maximal point
corresponding to the stable high-entropy solution and the other is a
minimal
point corresponding to the unstable low-entropy type. Usually a
minimal and a maximal points of a potential function denote a
stable and an
unstable solutions, respectively. However, since
we use the entropy of the system as a potential function,  the
correspondence
becomes reverse. When the mass of the black hole increases and
the solution point reaches to
the bifurcation set $B_V$, the maximal  and the minimal points
merge and
turn to
be a inflection point. Furthermore when the mass gets larger, i.e.,
the right hand
side of $B_V$, there is no extremal point and no black
hole solution. Catastrophe comes about with disappearance of a
maximal and
a minimal points.

 We plot a family of  the Skyrme black hole solutions in the 3-
dimensional
$M-B_{\rm H}-S$ space and its projections onto each 2-
dimensional plane.
This 3-dimensional picture  provides us clearer understanding and
some new results
(Fig. 7).
In catastrophe
theory, solutions are regarded as extremal points on the Whitney
surface,
$S=S(M, B_{\rm H})$, when the control parameter $M$ is fixed. At
the maximal
entropy in Fig. 7, the solution turns out to be an inflection point,
beyond
which there is no black hole solution. The projection curves onto
the $M-B_{\rm H}$
and $B_{\rm H}-S$ plane and the original 3-dimensional curve are
all
smooth, while  the projection onto the $M-S$ plane, and then that
onto the
$M-r_H$ plane shows a cusp structure.

A family of black hole solutions
connects two particle-like solutions with a smooth curve like a
``bridge",
though there is a ``pit fall" of the catastrophe on it.
One particle is ``stable" and the other is unstable.  This can be
understood through
the black hole solution with a catastrophe theory.
Such a relationship between the particle-like solutions
would not have been found out, if the associated black hole
solutions were
not investigated.

\subsection{Black Holes in a Heat Bath}

In thermodynamics, we often consider two complimentary
situations:
 one is an isolated adiabatic system and the other
 is an isothermal state in a heat bath.
For the former case, the entropy $S$ is the fundamental variable
and the
thermal equilibrium is realized at an entropy maximum,
 while for the latter case the Helmholtz free energy $F$ is the
fundamental
variable and the equilibrium is obtained at a minimum point of the
free energy.

In the black hole physics, we know that  black holes
have many analogous properties with the thermodynamical laws.
After Hawking discovered a thermal radiation from a black hole,
the thermodynamical interpretation of black hole physics
by this analogy turns to be more realistic. The black hole may be
regarded
as the a thermodynamical object with
the energy $M$, the entropy $S=A/4$ and the temperature $T=
\kappa/4 \pi$,
where $A$ and $\kappa$ are the area and the surface gravity of the
black hole,
respectively.
Although a black hole is not exactly  a thermodynamical
object, it may be very interesting to investigate
what will happen when a black hole is put in a ``heat reservoir".

Here we shall  consider an ideal
situation in which  a reservoir keeps the  temperature
of a black hole in a heat bath unchanged.
The thermal radiation around a black hole plays a role of heat
reservoir.
The radiation goes into the black hole or the Hawking radiation
comes
 out from the black hole to keep both temperatures same.
For example, the isolated Schwarzschild black hole
is stable, however the Schwarzschild black hole in a box,
which is filled by radiation,
is always unstable when the box is infinitely large\cite{Page}.
The reason why it is so may be understood by the fact that
 the specific heat is  negative.
As for the Reissner-Nordstr\"om
black hole,
there is a sign change of the specific heat at $Q=\sqrt{3} M/2$.
This sign change may correspond to a kind of phase transition\cite
{Davies},
i.e.,
we expect that the black hole in a heat bath becomes stable
when $Q>\sqrt{3}M/2$.

Although there is still a deep and unsolved problem about
the interpretation of such a sign change of
the specific heat and a stability in a ``heat bath"
\cite{Brown},
here we
will deal with a black hole just as a conventional thermodynamical
object.
Then the Helmholtz free energy
\begin{equation}
F=M-TS
\end{equation}
is a fundamental potential function.  The extremal point describes
an equilibrium state, i.e.,  the minimum and maximum correspond to
a stable and an unstable equilibrium states, respectively.
 Using the Helmholtz free energy  $F$, we shall
consider what will happen when the non-Abelian black holes are
put into a
heat bath.

\subsubsection{Type I non-Abelian black hole}

In the catastrophe theory of the dilatonic colored
black holes in a heat bath,
we adopt the temperature $T$, rather than mass $M$,
and the coupling constant
$\alpha$ as control parameters and   the Helmholtz free energy $F$
as a potential function instead of the entropy $S$
(see Table 3).
We plotted the equilibrium space in Fig. 8.
For the low coupling constant solutions,
which includes the colored black hole
as the case of $\alpha =0$,
two folds appear, while no fold is seen in the
case of high coupling constant.
This
configuration gives a cusp on the control plane through a
catastrophe map.
The bifurcation set is also shown in Fig.8, which
in fact has a cusp.
Hence we can conclude that
this type of stability belongs to a
cusp catastrophe.
As we can see from the behavior of the potential function shown in
Fig. 9,
a stable solution exists only in the  region ABC lying
between two curves AC and BC in Fig. 9, where two unstable
solutions exist as well. The stable solution is
characterized by the fact that the value of $\partial
B_{\rm H}/\partial T$ is negative in the equilibrium space, while
in the the most part of the equilibrium space where its value is
positive
the solutions are unstable.
This aspect is inverse of the normal cusp catastrophe, in which
stable
solutions exist  for any value of control parameters
and one unstable mode
appears only in a folded region.
Hence, strictly speaking, the system of Type I
black hole in a heat bath may be classified into a dual cusp
catastrophe.

This type of catastrophe reveals some interesting properties.
Suppose that there is a stable Type I black hole
which is expressed  by a point on the curve $e$ in Fig. 8.
When the temperature of the black hole either increases or
decreases,
the solution point reaches to  either ${\rm C_1}$ or ${\rm C_2}$,
beyond which there is no
stable solution.
Then the solution point may jump to the Schwarzschild black hole
along one of the solid arrows.
Though the final state are almost the same
whichever  the temperature increases or decreases,
 behaviors of the potential function are
 a little bit different as will be shown in below.

When the temperature of the black hole increases and the point
reaches
at C$_1$,
the maximal point $m_1$ and the minimal point
$m_2$  on the bifurcation set (the curve BC)
 merge, leaving  one maximal point $m_3$.
However, the black hole does not go to $m_3$ because it is
unstable. Instead the solution will jump to
more simple  Schwarzschild black hole.
When the
temperature decreases, and the solution point crosses the curve
AC, the minimal
point $m_2$ and the maximal point $m_3$ merge, leaving the
maximal point
$m_1$. The solution again jumps to the
Schwarzschild black hole.

As we mentioned, a catastrophe occurs on the bifurcation set.
Is there any physical
change which characterizes the bifurcation set in the present
model?
In the case of
Reissner-Nordstr\"om black hole in a heat bath, the bifurcation set
satisfies
$M=\frac{\sqrt{3}}{2} Q$, which is the point where
the specific heat changes its sign and
a kind of second order phase transition may take place.
For the present non-Abelian black holes, from our numerical
analysis,
we also expect that
the bifurcation set consists of the points where
the specific heat changes its sign.
This result is consistent with the fact that
the sign change occurs twice for the small  coupling constant
$\alpha$ while
no sign change occurs for the large values.

\subsubsection{Type II non-Abelian black hole}

The stability analysis of  Type II black hole in a heat bath
is more complicated.
We plotted the equilibrium space of the Skyrme black hole
and the bifurcation set on the control
plane (Fig. 10) as an example.
 The bifurcation set consists of two components, one of
which is the curve corresponding to a fold catastrophe and the
other is
the cusp shape corresponding to a cusp catastrophe.

We showed the behaviors of the potential function in Fig. 11.
Fixing
$\mu$, it is interesting to examine how the potential
function looks like as the temperature of the black hole changes.
There are qualitatively  four cases (a)$\sim$(d)
as shown by dotted lines in Fig. 11.
The shape change of the potential
function is shown  in Fig. 12.
For the  case (a), first one minimum $m_1$ and
one maximum $m_2$ exist for the low temperature. When the
temperature
increases (the solution point moves to right), the point
eventually reaches to the curve AB, where
an inflection point appears.
Beyond this point, since
there are  two minima and two maxima in the  region ABCD,
another maximum $m_3$ and minimum $m_4$ appears.
Increasing the temperature  further,
 the minimum $m_2$ and the maximum $m_3$ merge and turn to be
a
inflection point on the curve CD.
The maximum $m_1$ and the minimum $m_4$ are left.
Suppose that an initially  stable solution is at the minimum $m_2$.
As the  temperature increases,
it will disappear and jumps to the minimum $m_4$
beyond the curve CD. A catastrophe occurs.
Although the free energy at minimal point $m_4$ gets smaller than
that of
minimal point $m_2$ before the curve CD, the solution does not go
to
the point $m_4$, because the point  $m_2$ is locally stable.
Note that quantum effects break
this rule because of a quantum tunneling.

The case (b) after the curve EC is qualitatively not
 different from the case (a),
except that new phase appears before the curve EC.
For the case  (c),
the appearance of the maximum $m_3$ and the minimum $m_4$
 is accomplished at BF
earlier than that of $m_1$ and $m_2$, which occurs on the line BC.
The stable solution $m_4$, which  exists  initially at the low
temperature,
continues to exist and there happens no catastrophe even if the
temperature
increases.
Hence,  a stable black hole at the low temperature
can be  different
depending on the coupling constant $\alpha$.
When the temperature increases, some
 solutions
continue to exist while a catastrophe occurs for other solutions
and
the spacetime structure changes discontinuously. This is
interesting,
because we do not know
 whether our solution
is in the case (a) (or the case (b)) or in  the  case (c)
only from the black hole structure at the low
temperature.
The last case (d) is easy to understand because the shape of a
potential
 function changes only on the BF.

We found two types of elementary
catastrophe (a fold and a cusp types) from the bifurcation set.
Unless the number of
control parameters is more than two, however,
such a catastrophe can never happen at the same time
according to Thom's theorem\cite{Thom}.
Hence we expect  that this system may have
higher dimensional elementary catastrophe and
the bifurcation set in Fig. 10 is just a cross section of
a  higher bifurcation set.
It may be a swallow
tail catastrophe, a butterfly
catastrophe or else.


\vskip 2 cm
\setcounter{equation}{0}
\section{Concluding Remarks}

We have re-analyzed five known globally neutral
non-Abelian black holes and presented
a unified picture.
Those  are classified into two
types depending on whether the non-Abelian field is massless
(Type I) or massive (Type II).
The Type I non-Abelian black holes (the colored black hole and
the dilatonic colored black hole) have one particle-like solution
(the BM particle)
in zero entropy limit.  Although those black holes can have any
large mass,
the spacetime approaches to the Schwarzschild black hole
when its mass gets large. The specific heat changes
its sign twice
for the low coupling constant  or no changes for the high coupling
constant.
The Type II non-Abelian black holes have two types of solutions:
one
belongs to the ``stable" high-entropy branch and the other to the
 unstable low-entropy branch.
These two branches merge at a critical mass $M_{\rm C}$
 and provide a cusp there
in the  $M-S$ plane. There is no solution beyond the cusp
on which the solution has maximal mass and maximal entropy.
The high-entropy branch has no change of the sign of specific heat
and
resembles to the Schwarzschild de-Sitter black hole, while
the low-entropy branch changes the sign of specific heat a few
times
and is similar to
the Type I black hole. This is consistent with the fact that
the high-entropy branch
tends to a family of the Schwarzschild solutions and the
low-entropy branch converges to a family of the colored
black hole solutions as the mass of the
non-Abelian field decreases.

In order to analyze the stability of non-Abelian black hole, we
adopt a
catastrophe theoretic method. The Type II non-Abelian black hole
shows a fold
type of elementary catastrophe  and the Type I  black hole in a heat
bath
 has a cusp catastrophe. In case of the Type II  black hole
in a heat bath, we find a fold type and a cusp type structures in the
bifurcation set.
Since we may not have both in
one elementary catastrophe, this structures may be unified into
a higher dimensional elementary catastrophe.

Although a catastrophe theory may be a good tool to examine a
stability in nature,
we should give some critical comments as well.
First, when we discuss a stability change using a
catastrophe theory, we usually focus on some specific modes
though many modes
have to be investigated. Hence even if we find a stable branch
from the catastrophe theoretic analysis, it does not always mean
that such a system is completely stable.
 For example, the equilibrium space of the
sphaleron black hole is similar to that of the Skyrme black hole in
Fig. 5.
{}From its structure we might conclude that the lower (small $B_
{\rm H}$)
 solutions in the equilibrium
space are stable while the upper (large $B_{\rm H}$) solutions are
unstable.
 But this
is wrong for the sphaleron black holes
because both of solutions are unstable for the topological reason.
What is  true is that the lower solutions are stable for a certain
mode
while the upper solutions become unstable for the same mode.
Hence although the analysis is very easy,
we must be very careful to conclude about
a stability of a system by using a catastrophe theory.

Secondly, we have not succeeded to include the most stable
Schwarzschild
solution in our catastrophe theoretic analysis.
We can expect that
after a catastrophe occurs, the solution will jump to the
Schwarzschild black hole
branch.  But since the Schwarzschild solution does not
appear in the equilibrium space, we cannot expect which mass of
the black hole
will be found after the catastrophe. We have to solve the dynamical
equations.
However, since we have still the area theorem in the present
models\cite{Wal},
the area must increase. When the non-Abelian black hole collapses
 into the Schwarzschild
solution, a part of the energy of the YM or other fields
may be emitted into infinity.  In that case, the mass energy is not
conserved and the final Schwarzschild black hole becomes less
massive.
As a result, we can conclude that the non-Abelian black hole
will jump into the range between the same mass state, which
corresponds
to no emission of the non-Abelian fields,  and the same entropy
state,
which corresponds to a maximal emission of the fields.
If the entropy of black hole is not conserved, this catastrophic
jump
is a kind of first order transition.

\vskip 1cm

\noindent
-- Acknowledgments --

We would like to thank Gray W. Gibbons, Takuya Maki,
Osamu Kaburaki, Fjodor V.
Kusmartsev, and Ian Moss for useful discussions.
T. Tachizawa thanks for financial support by JSPS.
This work was supported
partially by the Grant-in-Aid for Scientific Research
Fund of the Ministry of
Education, Science and Culture  (No. 06302021 and No. 06640412),
by the Grant-in-Aid for JSPS Fellows (053769),
and by the Waseda University Grant for Special Research Projects.

\newpage
\begin{flushleft}
{ Figure Captions}
\end{flushleft}
\baselineskip = 24pt

\vskip 0.1cm
\baselineskip = 24pt
   \noindent
\parbox[t]{2cm}{\bf FIG 1:\\~}\ \
\parbox[t]{14cm}
{\baselineskip = 24pt
The mass-horizon radius diagrams for (a)
the Proca black hole with $\mu/g_{\rm C}m_{\rm P}=$ (i) 0.05, (ii)
0.10, and (iii) 0.15,
(b) the Skyrme black hole with
$\mu(=g_{\rm S} f_{\rm S})/g_{\rm S}m_{\rm P}=$ (i) 0.01, (ii)
0.02, and (iii) 0.03,
(c) the sphaleron black hole with $\mu(=g_{\rm C} \Phi_0)/g_{\rm
C}m_{\rm P}=$ (i) 0.1, (ii) 0.2, and
(iii) 0.3. $C$ is a cusp, where the black hole has a maximal entropy.
$C$ is a cusp, where the black hole has a maximal entropy.
Beyond its entropy there is no non-Abelian black hole. The
Schwarzschild black
hole (the dot-dashed line) and the colored black hole (the dotted
line)
are also shown as references.
The high-entropy (large horizon radius) branch is ``stable",
while the low-entropy (small horizon radius)
branch is unstable.}\\[1em]
\noindent
\parbox[t]{2cm}{\bf FIG 2:\\~}\ \
\parbox[t]{14cm}
{\baselineskip = 24pt
The distributions of energy density for (a) the colored black hole
(Type I),
(b) low-entropy branch, and (c) the high-entropy branch of the
Proca black hole (Type II) with $\mu /g_{\rm C}m_{\rm P}=0.05$.
The horizon
radius of each black hole is $0.01g_{\rm C}/l_{\rm P}$. We plot the
enerty
densities $\rho_{F^2}$ from the kinetic term
 and $\rho_{A^2}$ from the mass term separately. At
 the Compton Wave length of the massive YM field
$(\sim 1/\mu )$, which marked by an arrow,
$\rho_{A^2}$ is still dominant in the high-entropy branch while
$\rho_{F^2}$ becomes dominant in the low-entropy branch just the
same
as the colored black hole.
}\\[1em]
 \noindent
\parbox[t]{2cm}{\bf FIG 3:\\~}\ \
\parbox[t]{14cm}
{\baselineskip = 24pt
The mass-temperature relations for (a)
the Proca black hole with $\mu/g_{\rm C}m_{\rm P}=$ (i) 0.05, (ii)
0.10, and (iii) 0.15,
(b) the Skyrme black hole with
$\mu/g_{\rm S}m_{\rm P}=$ (i) 0.01, (ii) 0.02, and (iii) 0.03,
(c) the sphaleron black hole with $\mu/g_{\rm C}m_{\rm P}=$ (i)
0.1, (ii) 0.2, and
(iii) 0.3.  The specific heat in the high-entropy branch
is always negative. While, that in the low-entropy
branch changes its sign three times for small values of $\mu$
and once for large values of $\mu$.
}\\[1em]
   \noindent
\parbox[t]{2cm}{\bf FIG 4:\\~}\ \
\parbox[t]{14cm}
{\baselineskip = 24pt
The field strength at the horizon $B_{\rm H}$ with respect to
$M$ for the colored black hole (Type I)  and the
Proca black hole with $\mu=0.05g_{\rm C}m_{\rm P}$
(Type II). For the Type I and the low-entropy branch of the Type II,
we find rather large value of $B_{\rm H}$, which means that
the black hole is locally charged near the horizon.  When the black
hole gets large
its value decreases because the spacetime approaches to the
Schwarzschild solution.  For the high-entropy branch of the Type II,
however,
the value of $B_{\rm H}$ is very small.  Such a black hole is almost
neutral everywhere.
}\\[1em]
   \noindent
\parbox[t]{2cm}{\bf FIG 5:\\~}\ \
\parbox[t]{14cm}
{\baselineskip = 24pt
The equilibrium space of the Skyrme black hole. Upper side and
lower side
of the equilibrium space correspond to the unstable and stable
solutions
respectively. When  the black hole mass gets large (along the curve
$e$),
the stable and the unstable solutions merge at a critical mass
scale.
 A catastrophe occurs at this critical point and
the Skyrme
black hole will jump to the  Schwarzschild black hole (along the
solid arrow).
On the control plane,  we draw the bifurcation set $B_V$, which
shows a fold
catastrophe.
To  make it
easy to see,  we have moved down the level of the control plane
lower than the original position.
}\\[1em]
   \noindent
\parbox[t]{2cm}{\bf FIG 6:\\~\\~}\ \
\parbox[t]{14cm}
{\baselineskip = 24pt
The potential function $S$ (entropy) of the Skyrme black hole.
A maximal point and a minimal point exist in the left hand side of
the bifurcation
set $B_V$, while in the right hand side there is no extremal point.
}\\[1em]
   \noindent
\parbox[t]{2cm}{\bf FIG 7:\\~\\~}\ \
\parbox[t]{14cm}
{\baselineskip = 24pt
The solution curve in the three dimensional space of $(M, B_{\rm
H}, S)$
and its projections onto each two dimensional plane for the Skyrme
black hole
with $\mu=0.02 g_{\rm S} m_{\rm P}$. The cusp $C$ in $M-S$ plane
is a critical point
for stability. For the fixed control parameter $M$, there are two
solutions at
extremal points on the Whitney surface: the maximal one is stable,
but the minimal one is unstable. Beyond the critical point $C$,
there
is no extremal point, i.e., no non-Abelian black hole.
}\\[1em]
   \noindent
\parbox[t]{2cm}{\bf FIG 8:\\~\\~}\ \
\parbox[t]{14cm}
{\baselineskip = 24pt
The equilibrium space and the bifurcation set of the dilatonic
colored black
hole in a heat bath. From the shape of the bifurcation set, we
classify
this system into a cusp catastrophe.
}\\[1em]
   \noindent
\parbox[t]{2cm}{\bf FIG 9:\\~\\~}\ \
\parbox[t]{14cm}
{\baselineskip = 24pt
The potential function $F$ (Helmholtz free energy)
of the dilatonic colored black
hole in a heat bath. In the interior of the region ABC, there are
three extremal
points $m_1, m_2,$ and $ m_3$, one of which ($m_2$) is stable and
others
($m_1, m_3$) are unstable.
 In the other region, however,
only one unstable solution. Such a configuration is not seen in a
usual cusp catastrophe but may be regarded as  a dual cusp
catastrophe.
}\\[1em]
   \noindent
\parbox[t]{2cm}{\bf FIG 10:\\~\\~}\ \
\parbox[t]{14cm}
{\baselineskip = 24pt
The equilibrium space and the bifurcation set of the Skyrme black
hole in a heat bath. The bifurcation set consists
of a cusp structure ECD and a smooth curve ABF.
The appearance of both a smooth curve and a cusp
at same time means that this system is be classified
into one elementary catastrophe with two control parameters but
may into
higher type (a swallow's tail or a butterfly whose number of the
control
parameters is higher
than three).
}\\[1em]
   \noindent
\parbox[t]{2cm}{\bf FIG 11:\\~\\~}\ \
\parbox[t]{14cm}
{\baselineskip = 24pt
The potential function $F$ (Helmholtz free energy)
of the Skyrme black
hole in a heat bath. In the   region ABCD, there are two maxima
 and two minima points ($m_1 \sim  m_4$). The four lines labeled
(a)$\sim$(d) are used
in the discussion about the evolution of the black holes.
Each line correspond to different values of the mass $\mu$.
}\\[1em]
   \noindent
\parbox[t]{2cm}{\bf FIG 12:\\~\\~}\ \
\parbox[t]{14cm}
{\baselineskip = 24pt
The behaviors of the potential function of the Skyrme black hole
in a heat bath. There are four cases labeled (a)$\sim$(d) which
correspond to
the line labeled (a)$\sim$(d) in Fig.11.  The
characters at the bottom of  several figures, e.g. AB, CD,
 mean that the figures are those on the curves EBC, CD, or ABF of
the
bifurcation set in Fig. 11.
}

\newpage
\begin{flushleft}
{Table Captions}
\end{flushleft}
\baselineskip = 24pt

\vskip 0.1cm
\baselineskip = 24pt
   \noindent
\parbox[t]{3cm}{\bf TABLE 1:\\~}\ \
\parbox[t]{13cm}
{\baselineskip = 24pt
The models we re-analyzed and the names of the non-Abelian
black holes and non-trivial particles
We classify them into two types;
Type I (models with massless non-Abelian field) and Type II
(models with massive non-Abelian field).
}\\[1em]
   \noindent
\parbox[t]{3cm}{\bf TABLE 2:\\~}\ \
\parbox[t]{13cm}
{\baselineskip = 24pt
The properties of Type I and II  black holes.
``small" or ``low" in Type I means that its value is similar to
 small  or low in Type II.
See text about the
meaning of ``stable" for the sphaleron black hole.
$C_\#$ denotes how many times the sign of the specific
heat changes in the branch.
}\\[1em]
   \noindent
\parbox[t]{3cm}{\bf TABLE 3:\\~}\ \
\parbox[t]{13cm}
{\baselineskip = 24pt
The control parameters, the state variable  and the potential
function
of  Type I and II black holes.
The  type  of elementary catastrophe for each system is also given.
}\\[1em]
   \noindent


\begin{thebibliography}{99}
\bibitem{comment} However, if we consider the case that the
coupling constants
                 change with the temperature or the energy scale as
running coupling constants,
                 we may find something different in the history of the
universe from
                 the discussion here.
\end{thebibliography}
\end{document}